\documentclass[aps,preprint,showpacs,preprintnumbers,amsmath,amssymb]{revtex4}
\usepackage{amsmath,mathrsfs,amsbsy,color,graphicx,bm,amsthm,amsfonts}
\usepackage{units}
\usepackage{bbm}
\usepackage{times}
\usepackage{dcolumn}
\usepackage{mathrsfs}
\usepackage{amsmath,amssymb,epsfig}
\usepackage{amsmath}
\usepackage{graphicx}
\newcommand{\udots}{\mathinner{\mskip1mu\raise1pt\vbox{\kern7pt\hbox{.}}
\mskip2mu\raise4pt\hbox{.}\mskip2mu\raise7pt\hbox{.}\mskip1mu}}
\begin{document}
\title{Entropic uncertainty and coherence in Einstein-Gauss-Bonnet gravity}
\author{ Wen-Mei Li$^1$, Jianbo Lu$^1$\footnote{Email: lvjianbo819@163.com}, Shu-Min Wu$^1$\footnote{ Email: smwu@lnnu.edu.cn}}
\affiliation{$^1$ Department of Physics, Liaoning Normal University, Dalian 116029, China}


\begin{abstract}
We investigate  tripartite quantum-memory-assisted entropic uncertain and quantum coherence for GHZ and W states of a fermionic field in the background of a spherically symmetric black hole of Einstein-Gauss-Bonnet (EGB) gravity. Two distinct scenarios are analyzed: (i) the quantum memories (held by Bob and Charlie) are near the horizon while the measured particle (Alice) remains in the flat region, and (ii) the reverse configuration.  Dimensional dependence is observed: in $d>5$ dimensions, the measurement  uncertainty decreases monotonically with increasing horizon radius, while coherence increases; in
$d=5$, both quantities exhibit non-monotonic behavior due to distinctive thermodynamic properties.
Furthermore, comparative analysis reveals that the W state exhibits higher robustness in preserving coherence, whereas the GHZ state shows greater resistance to measurement  uncertainty increase induced by Hawking radiation. Notably, the two scenarios yield qualitatively distinct behaviors: quantum coherence is consistently lower in Scenario 1 (quantum memory near horizon) than in Scenario 2 (measured particle near horizon), irrespective of the quantum state. For measurement uncertainty, the W state displays lower uncertainty in Scenario 1, while the GHZ state exhibits the opposite trend, with higher measurement  uncertainty in Scenario 1. These results indicate that the characteristics of different quantum resources provide important insights into the selection and optimization of quantum states for information processing in curved spacetime.
\end{abstract}

\vspace*{0.5cm}
 \pacs{04.70.Dy, 03.65.Ud,04.62.+v }
\maketitle
\section{Introduction}
The concept of higher-dimensional spacetime, motivated by heterotic string theory and braneworld models~\cite{L3,L4,L5,L6,LL7}, extends general relativity beyond its conventional (3+1)-dimensional framework by introducing additional compact or extended dimensions. In the braneworld scenario, matter and gauge fields are confined to a (3+1)-dimensional brane, while gravity propagates in the higher-dimensional bulk, necessitating a generalized theory of gravity. Among such extensions, Lovelock gravity provides a natural and ghost-free framework~\cite{L10,L11,L14,LL3}, with its second-order term yielding EGB gravity~\cite{L15,L16}, which includes the cosmological constant, Einstein-Hilbert, and Gauss-Bonnet terms. The Gauss-Bonnet correction offers a nontrivial extension of Einstein gravity, leading to rich implications for black hole physics, thermodynamics, and holography. As four-dimensional general relativity remains incompatible with a complete quantum theory, exploring the quantum properties of higher-dimensional black holes in EGB gravity provides a promising route toward understanding the interplay between gravity and quantum effects~\cite{L18,L19,L20,LL20}.

Relativistic quantum information is an interdisciplinary field at the intersection of quantum information theory, quantum field theory, and general relativity \cite{R1,RR1,R2,R3,R4,RR2,R5,R6,R7,R8,R9,R10,R11,R12,R13,R14,R15,R16,R17,R18,R19,R20,R21,
R22,R23,R24,R25,R26,R27,R28,R29,R30,R31,R32,R33,R34,R35,R36,R37,R38,R39,R40,R41,R42,R43,
ENM5,R44,R45,R46,R47,R48,R49,R50,RR51,R51,R52,ENM1,ENM2,ENM4,ENM6,ENM7,ENM8,ENM9,AVM1,AVM2}. It has seen rapid theoretical and experimental development, offering novel insights into quantum phenomena in curved and Rindler spacetimes. In particular, quantum resources have been extensively explored in
four-dimensional spacetime such as Schwarzschild black holes, Garfinkle-Horowitz-Strominger (GHS) dilaton black holes, and uniformly accelerated (noninertial) frames \cite{R1,RR1,R2,R3,R4,RR2,R5,R6,R7,R8,R9,R10,R11,R12,R13,R14,R15,R16,R17,R18,R19,R20,R21,R22,R23,R24,R25,R26,R27,R28,R29,R30,R31,R32,R33,R34}.
Despite substantial progress in related areas, the fundamental characteristics and distinctions between the entropic uncertainty relation (EUR) and quantum coherence in the context of EGB gravity remain unclear, particularly regarding their behavior for different quantum states. Since different quantum states may require different types of quantum resources, a thorough understanding of these features is essential for efficiently performing relativistic quantum information processing tasks. This gap in understanding serves as the key motivation for the present study.  The uncertainty principle imposes fundamental limits on the simultaneous predictability of two
non-commuting observables. To more accurately quantify this limitation, entropic formulations of the uncertainty principle were developed and subsequently extended to quantum-memory-assisted EUR, which reveal that quantum correlations can effectively reduce uncertainty \cite{R53,R54}.  In parallel, quantum coherence viewed as a resource arising from basis-dependent superposition has also been found to contribute directly to entropic uncertainty \cite{R55}. Recently, the connection between quantum coherence and EUR has attracted attention and achieved some progress, but its underlying mechanism remains unclear \cite{R54,R56}.
In this work, we focus on GHZ and W states as probes to systematically analyze the interplay between EUR and quantum coherence in the background of EGB gravity. By contrasting their distinct quantum informational responses to the gravitational environment, we aim to reveal how spacetime geometry influences fundamental quantum properties, thereby offering new insights into the physical implications of EGB gravity.

In this work, we consider a tripartite system consisting of Alice, Bob, and Charlie, who initially share GHZ and W states in the asymptotically flat region of a spherically symmetric black hole in EGB gravity. Two distinct scenarios are analyzed: in Scenario 1, Bob and Charlie, who hold the quantum memories $B$ and $C$, approach the event horizon while the measured particle held by Alice remains in the asymptotically flat region; in Scenario 2, Alice hovers near the event horizon, whereas Bob and Charlie remain far away in flat spacetime. Within these settings, we will investigate the behavior of measurement uncertainty and quantum coherence for both GHZ and W states, highlighting how gravitational effects in EGB spacetime influence quantum resources.
For both cases, we derive analytical expressions for measurement uncertainty and coherence, and identify two noteworthy features: (i) in Scenario 1, the influence of EGB gravity on the measurement uncertainty is not always stronger than in Scenario 2, but its effect on quantum coherence is consistently more pronounced; (ii) the GHZ state demonstrates greater resistance to gravitational effects in terms of measurement uncertainty, whereas the W state exhibits superior robustness in maintaining quantum coherence under the same conditions.  Our results reveal that the GHZ and W states exhibit complementary advantages in EGB gravity. The GHZ state is more resilient to gravitational effects on measurement uncertainty, whereas the W state preserves quantum coherence more effectively, thereby enriching our understanding of relativistic quantum information in higher-dimensional gravitational backgrounds.

The structure of the paper is as follows.  Sec. II presents the quantization of the fermionic field in EGB gravity. Sec. III outlines the EUR and the measures of quantum coherence. Sec. IV investigates the behavior of measurement uncertainty and coherence for the fermionic field in EGB spacetime. Finally, Sec. V provides concluding remarks.

\section{Quantization of Dirac field in EGB spacetime \label{GSCDGE}}

The field equation of EGB gravity reads
\begin{eqnarray}\label{w5}
G_{\mu \nu}^{(\mathrm{E})}+\Lambda g_{\mu \nu}+\alpha G_{\mu \nu}^{(\mathrm{GB})}=T_{\mu \nu},
\end{eqnarray}
where $G_{\mu \nu}^{(\mathrm{E})}$ denotes the Einstein tensor, $\alpha$ is the Gauss-Bonnet coupling constant (we take as $\alpha >0$) \cite{M7,L7}, $G_{\mu \nu}^{(\mathrm{GB})}$ is the second-order Lovelock tensor or Gauss-Bonnet tensor, and $T_{\mu \nu}$ is the energy-momentum tensor of matter. $G_{\mu \nu}^{(\mathrm{GB})}$ is defined in terms of the curvature tensor $R_{\mu \sigma \kappa \tau}$ as
\begin{eqnarray}\label{w6}
G_{\mu \nu}^{(\mathrm{GB})}&&=2\left(R_{\mu \sigma \kappa \tau} R_{\nu}{ }^{\sigma \kappa \tau}-2 R_{\mu \rho \nu \sigma} R^{\rho \sigma}-2 R_{\mu \sigma} R^{\sigma}{ }_{\nu}+R R_{\mu \nu}\right) \notag\\
&&-\frac{1}{2}\left(R_{\mu \nu \sigma \kappa} R^{\mu \nu \sigma \kappa}-4 R_{\mu \nu} R^{\mu \nu}+R^{2}\right) g_{\mu \nu}.
\end{eqnarray}
One considers a $d$-dimensional ($d\ge 5$) static spherically symmetric spacetime with the metric
\begin{eqnarray}\label{w7}
ds^{2}=-f(r) d t^{2}+\frac{1}{f(r)} d r^{2}+r^{2} d \Omega_{d-2}^{2},	
\end{eqnarray}
where $f(r)$ is the metric function  and $r^{2} \rm{d} \Omega_{d-2}^{2}$ denotes the metric of a $(d-2)$-dimensional subspace \cite{M10,M8,M9}. It can be proved that this metric describes a black hole solution of the field Eq.(\ref{w5}) with $\Lambda=0$, provided that
\begin{eqnarray}\label{w8}
f(r)=k+\frac{r^{2}}{2(d-3)(d-4) \alpha}\left(1 \pm \sqrt{1+\frac{4(d-3)(d-4) \alpha m}{r^{(d-1)}}}\right),
\end{eqnarray}
where $k$ is the curvature of the $(d-2)$-dimensional subspace and $m$ denotes the mass of the black hole \cite{M10}. For the special case of $d=5$,  the function takes a particular form as
\begin{eqnarray}\label{w9}	
f(r)=k+\frac{r^{2}}{4 \alpha} \pm \sqrt{\frac{r^{4}}{16 \alpha^{2}}+\left(|k|+\frac{m}{2 \alpha}\right)},
\end{eqnarray}
which has a geometrical mass $m+2\alpha|k|$. Since our research mainly focuses on the asymptotically flat solutions, we set $k=1$, and for Eqs.(\ref{w8}) and (\ref{w9}), we adopt the negative sign. In the limit  $ \alpha\rightarrow 0$, the $f(r)$ reduces to the  $d$-dimensional Schwarzschild solution, as required by Lovelock gravity. Obviously, the horizon radius $r_{h}$ is determined by the largest root of $f(r)=0$; e.g. for $d=5$, one finds $r_{h}=\sqrt{m}$. It is worth stressing that the thermodynamic behavior in  $d=5$ differs qualitatively from higher dimensions. For $d>5$, the Hawking temperature diverges as the black hole mass vanishes, whereas in $d=5$ it tends to zero. Consequently, five-dimensional Gauss-Bonnet black holes exhibit a positive heat capacity and are locally thermodynamically stable, in sharp contrast to their
higher-dimensional counterparts. A detailed discussion of this phenomenon can be found in references \cite{M8, M2}.

By an appropriate coordinate transformation, one can introduce the Kruskal coordinates.
In terms of the null coordinates  $(u,v)$, the metric reads
\begin{eqnarray}\label{w10}
&&d s^{2}=-f(r) d u d v+r^{2} d \Omega_{d-2}^{2}, \notag\\
&&u=t-r_{*}, v =t+r_{*},	
\end{eqnarray}
where the Regge-Wheeler tortoise coordinate is defined as $r_{*}=\int dr/f(r)$. Near the horizon $r_{h}$, it can be expanded as
\begin{eqnarray}\label{w11}	
r_{*} & \approx \Gamma \ln \left(r-r_{h}\right)+G\left(r-r_{h}\right),
\end{eqnarray}
where $G\left(r-r_{h}\right)$ is a nonsingular function at $r_{h}$.
The coefficient $\Gamma$ is identified with the inverse Hawking temperature, $\Gamma^{-1}=T$. Geometrically, the temperature is determined by the surface gravity
\begin{eqnarray}\label{w12}
T & =2 \pi\left(-\frac{1}{2} \nabla_{\mu} \xi_{\nu} \nabla^{\mu} \xi^{\nu}\right)^{-(1 / 2)},	
\end{eqnarray}
which for the metric of Eq.(\ref{w7}) reduces to \cite{M9}
\begin{eqnarray}\label{w13}
T & =\left.\frac{1}{4 \pi} \frac{d f(r)}{d r}\right|_{r_{h}}	.
\end{eqnarray}
Evaluating this expression for  Eqs.(\ref{w8}) and (\ref{w9}), one finds
\begin{eqnarray}\label{w14}
T=\frac{1}{4 \pi} \frac{\alpha d^{3}-12 \alpha d^{2}+47 \alpha d+d r_{h}^{2}-60 \alpha-3 r_{h}^{2}}{r_{h}\left(2 \alpha d^{2}-14 \alpha d+24 \alpha+r_{h}^{2}\right)},
\end{eqnarray}
while in the special case $d=5$,  it reduces to
\begin{eqnarray}\label{w15}	
T=\frac{1}{2 \pi} \frac{r_{h}}{\left(4 \alpha+r_{h}^{2}\right)}.
\end{eqnarray}
For an analytic extension of the metric, the Kruskal coordinates are defined as \cite{R5}
\begin{eqnarray}\label{w16}
U\propto \pm e^{-uT} ,V \propto \mp e^{-vT}.
\end{eqnarray}

We consider the massless Dirac field equation, which can be expressed as
\begin{eqnarray}\label{w17}
[\gamma^{a}e_{a}^{u}(\partial_{u}+\Gamma_{u})]\Phi=0,
\end{eqnarray}
where $\gamma^{a}$ are the Dirac gamma matrices, the four-vectors $e_{a}^{u}$ is the inverse of the tetrad $e_{u}^{a}$, and $\Gamma_{u}$ are the spin connection coefficients \cite{QTU1}. Eq.(\ref{w17}) in the EGB spacetime can be specifically written as
\begin{eqnarray}\label{w18}
&-&\frac{\gamma_{0}}{\sqrt{f(r)}}\frac{\partial\Phi}{\partial t}+\gamma_{1}\sqrt{f(r)}\bigg[\frac{\partial}{\partial r}+\frac{1}{2}\bigg(\frac{f'(r)}{2f(r)}+\frac{2}{r}\bigg)\bigg]\Phi\notag\\
&+&\frac{\gamma_{2}}{r}\bigg(\frac{\partial}{\partial\theta}+\frac{\cot \theta }{2}\bigg)\Phi+\frac{\gamma_{3}}{r\sin\theta}\frac{\partial\Phi}{\partial\varphi}=0.
\end{eqnarray}
Solving the Dirac equation near the event horizon, we obtain a set of positive (fermions) frequency outgoing solutions for the inside and outside regions of the event horizon as
\begin{eqnarray}\label{w20}	
&&\Phi_{\mathrm{{\bf k}}, \mathrm{II}}^{+} \sim \phi(r)e^{i \omega u},\notag\\
&&\Phi_{\mathrm{{\bf k}}, \mathrm{I}}^{+} \sim \phi(r)e^{-i \omega u},
\end{eqnarray}
where $\phi(r)$ is a four-component Dirac spinor and $\mathrm{{\bf k}}$ is the wave vector labeling the modes hereafter. The outside (region I) and inside (region II) regions are causally disconnected.
The Dirac field can then be expanded as
\begin{eqnarray}\label{w21}	
\Phi=\int d \mathrm{{\bf k}}\left[a_{\mathrm{{\bf k}}}^{\mathrm{II}} \Phi_{\mathrm{{\bf k}}, \mathrm{II}}^{+}+b_{\mathrm{{\bf k}}}^{\mathrm{II}\dagger} \Phi_{\mathrm{{\bf k}}, \mathrm{II}}^{-}+a_{\mathrm{{\bf k}}}^{\mathrm{I}} \Phi_{\mathrm{{\bf k}}, \mathrm{I}}^{+}+b_{\mathrm{{\bf k}}}^{\mathrm{I}\dagger} \Phi_{\mathrm{{\bf k}}, \mathrm{I}}^{-}\right],
\end{eqnarray}
where  $a_{\mathrm{{\bf k}}}^{\mathrm{II}}$ and $b_{\mathrm{{\bf k}}}^{\mathrm{II} \dagger}$ are the fermion annihilation and antifermion creation operators for the quantum field in the interior of the event horizon, and $a_{\mathrm{{\bf k}}}^{\mathrm{I}}$ and $b_{\mathrm{{\bf k}}}^{\mathrm{I} \dagger}$ are the fermion annihilation and antifermion creation operators for the quantum field in the exterior region acting on the quantum state, respectively. They satisfy the canonical anticommutation relations
\begin{eqnarray}\nonumber
\left\{a_{\mathbf{k}}^{\mathrm{II}},a_{\mathbf{k^{'}}}^{\mathrm{II}\dagger}\right\} =\left\{a_{\mathbf{k}}^{\mathrm{I}},a_{\mathbf{k^{'}}}^{\mathrm{I}\dagger}\right\}=\left\{b_{\mathbf{k}}^{\mathrm{II}},b_{\mathbf{k^{'}}}^{\mathrm{II}\dagger}\right\}=\left\{b_{\mathbf{k}}^{\mathrm{I}},b_{\mathbf{k^{'}}}^{\mathrm{I}\dagger}\right\}=\delta_{\mathbf{kk^{'}}}.
\end{eqnarray}
One can define the EGB vacuum $a_{\mathbf{k}}^{\mathrm{II}}\left |0\right \rangle_{\mathrm{II}}=a_{\mathbf{k}}^{\mathrm{I}}\left |0\right \rangle_{\mathrm{I}}=0$. Therefore, the modes $\Phi_{\mathbf{k},\sigma}^{\pm}$  with $\sigma=(\mathrm{II},\mathrm{I})$ are called  EGB modes.

Since the solutions of Eq.(\ref{w20}) cannot be analytically continued from region I to region II, it is convenient to express them in the Kruskal coordinates \cite{QTU2}.
Using Eqs.(\ref{w16}) and (\ref{w20}), one obtains the analytic mode solutions valid in the whole spacetime
\begin{eqnarray}\label{w23}
&&\Psi_{\mathbf{k},\mathrm{I}}^{+}=e^{ \omega /4 T} \Phi_{\mathbf{k},\mathrm{I}}^{+}+e^{-\omega / 4 T} \Phi_{\mathbf{k}, \mathrm{II}}^{-},\notag\\	
&&\Psi_{\mathbf{k},\mathrm{II}}^{
+}=e^{- \omega / 4 T} \Phi_{\mathbf{k},\mathrm{I}}^{-}+e^{ \omega / 4 T} \Phi_{\mathbf{k},\mathrm{II}}^{+}.
\end{eqnarray}
Therefore, the Dirac field can then be expanded in terms of Kruskal modes as
\begin{eqnarray}\label{w22}
\Phi= & \int d \mathbf{k}\left[\hat{c}_{\mathbf{k}}^{\mathrm{II}} \Psi_{\mathbf{k}, \mathrm{II}}^{+}\right. \left.+\hat{d}_{\mathbf{k}}^{\mathrm{II} \dagger} \Psi_{\mathbf{k},\mathrm{II}}^{-}+\hat{c}_{\mathbf{k}}^{\mathrm{I}} \Psi_{\mathbf{k}, \mathrm{I}}^{+}\right. \left.+\hat{d}_{\mathbf{k}}^{\mathrm{I} \dagger} \Psi_{\mathbf{k},\mathrm{I}}^{-}\right],
\end{eqnarray}
where $\hat{c}_{\mathbf{k}}^{\sigma}$ and $\hat{d}_{\mathbf{k}}^{\sigma\dagger}$ with $\sigma=(\mathrm{II},\mathrm{I})$ are the fermion annihilation and antifermion creation operators acting on the Kruskal vacuum.
Eqs.(\ref{w21}) and (\ref{w22}) represent two different decompositions of the same Dirac field, in terms of EGB and Kruskal modes, respectively, which are related by the Bogoliubov transformations
\begin{eqnarray}\label{w24}	
\begin{aligned}
\hat{c}_{\mathbf{k}}^{\mathrm{I}} & =\cos\zeta\hat{a}_{\mathbf{k}}^{\mathrm{I}}-\sin \zeta\hat{b}_{\mathbf{k}}^{\mathrm{II} \dagger}, \\
\hat{c}_{\mathbf{k}}^{\mathrm{I} \dagger} & =\cos\zeta\hat{a}_{\mathbf{k}}^{\mathrm{I} \dagger}-\sin \zeta \hat{b}_{\mathbf{k}}^{\mathrm{II}},
\end{aligned}
\end{eqnarray}
in which $\tan\zeta =e^{-\pi \omega /T}$. In this representation, the Kruskal vacuum and the first excited state  in EGB gravity are given by
\begin{eqnarray}\label{w29}
	|0\rangle_{K}=\cos\zeta|0\rangle_{\mathrm{I}} \otimes|0\rangle_{\mathrm{II}}+\sin \zeta|1\rangle_{\mathrm{I}} \otimes|1\rangle_{\mathrm{II}},
\end{eqnarray}
\begin{eqnarray}\label{w30}
	|1\rangle_{K}=|1\rangle_{\mathrm{I}}\otimes|0\rangle_{\mathrm{II}},
\end{eqnarray}
where $\{|n\rangle_{\text {I}}\}$ and $\{|n\rangle_{\text {II}}\}$ denote the fermionic number states in the exterior region and the corresponding antifermionic states in the interior region of the event horizon, respectively.

\section{EUR and quantum coherence \label{GSCDGE}}
The Heisenberg uncertainty principle, proposed by Heisenberg in 1927 \cite{M19}, is a cornerstone of quantum mechanics, stating that the position and momentum of a particle cannot be simultaneously measured with arbitrary precision. Kennard  and Robertson  formulated this principle as a standard deviation relation $\bigtriangleup R \bigtriangleup S\ge \left | \left \langle \left [ R,S \right ]  \right \rangle  \right |/2$, which is applicable to any incompatible observable measurements $R$ and $S$ and $\left [ R,S \right ]=RS-SR$ represents the commutator \cite{M20,M21,M22}. However, the variance-based formulation is not optimal since the bound depends on the quantum state. In information theory, entropy provides a more suitable quantifier of uncertainty. In this framework, the uncertainty principle can be expressed in terms of Shannon entropy \cite{M27,M28,M29}
 \begin{eqnarray}\label{w1}
H(R)+H(S)\ge -\log_{2}{c},
\end{eqnarray}
where $H(P)=-\sum _{k}p_{k}\log_{2}{p_{k}}$ is the Shannon entropy of measurement outcomes for observable $P\in \left \{ R,S \right \}$,
in which $p_{k}$ is the probability of obtaining measurement outcome of $k$, and the term $c$ measures the complementarity between the observables and is defined as $c=\max_{X,Z} \left | \left \langle x_{i}   | z_{i}  \right \rangle  \right |^{2} $, with $X=\left \{ \left | x_{i}  \right \rangle  \right \}$ and $Z=\left \{ \left | z_{i}  \right \rangle  \right \}$ being the eigenbasis of the observables $X$ and $Z$, respectively.

This relation has been generalized in various directions. In particular, Boileau and Renes introduced the quantum-memory-assisted entropic uncertainty relation (QMA-EUR) for bipartite and tripartite systems \cite{M32,M33,M30,M31,M34,M35}. For the tripartite case, the relation takes the form
\begin{eqnarray}\label{w2}
 U \equiv S(X |B)+  S(Z |C)\ge -\log_{2}{c},
\end{eqnarray}
where $S(X |B)=S(\rho _{X B} )-S\left ( \rho _{B}  \right )$ is the conditional entropy of state $\rho _{X B}$, and
\begin{eqnarray}\label{w3}
\rho _{X B}=\sum _{i} (\left|x_{i}\right\rangle_{A}\left\langle x_{i}\right|\otimes \bm{\mathrm{I}}_{B})\rho _{AB}(\left|x_{i}\right\rangle_{A}\left\langle x_{i}\right|\otimes \bm{\mathrm{I}}_{B}),
\end{eqnarray}
and
\begin{eqnarray}\label{w4}
\rho _{Z C}=\sum _{j} (\left|z_{j}\right\rangle_{A}\left\langle z_{j}\right|\otimes \bm{\mathrm{I}}_{C})\rho _{AC}(\left|z_{j}\right\rangle_{A}\left\langle z_{j}\right|\otimes \bm{\mathrm{I}}_{C}),
\end{eqnarray}
respectively, where $\bm{\mathrm{I}}_{B(C)}$ is the identity operator. $S(X|B)$ is used to measure Bob's uncertainty regarding Alice's measurement result for the observable $X$. In simple terms, the uncertain relationship can be described by an uncertain game. Suppose there are three friends, namely Alice, Bob, and Charlie. Initially, they share a quantum state $\rho _{ABC}$. Next, Alice selects one of the two measurements to execute the command, either $X$ or $Z$, and informs her choice to Bob and Charlie who respectively hold the quantum memory $B$ and $C$. If she chooses $X$, Bob's task is to predict Alice's measurement result; if she chooses $Z$, then it is Charlie's task to predict Alice's measurement result. In addition, the QMA-EUR has attracted considerable attention due to its wide range of applications, including quantum metrology and entanglement detection \cite{R53}. Experimentally, significant progress has been made in verifying and applying QMA-EUR across different platforms \cite{M46,MM51,M48,MM52,M51}.

It has been established that quantum coherence, as a basis-dependent resource originating from the superposition principle, directly contributes to the EUR. Coherence is a fundamental feature of quantum mechanics and plays a crucial role as a prerequisite for the generation of quantum entanglement. To quantify coherence, we employ two standard measures:
the $l_{1}$-norm of coherence and the relative entropy of coherence (REC) \cite{M37}. The
 $l_{1}$-norm of coherence is defined as the sum of the absolute values of the off-diagonal elements of the density matrix $\rho_{ABC}$
\begin{eqnarray}\label{w61}
C_{l_{1}}(\rho_{AB})=\sum_{i \neq j}\left|\rho_{i, j}\right|,
\end{eqnarray}
while the REC is given by the difference between the von Neumann entropy of the diagonalized state $S(\rho_{diag})$ and that of the full state  $\rho_{AB}$
\begin{eqnarray}\label{w62}
C_{REC}(\rho_{AB})=S(\rho_{AB_{diag}})-S(\rho_{AB}).
\end{eqnarray}
Since coherence determines the degree of quantum superposition available in the system, it naturally influences the uncertainty bound in the EUR with quantum memory, thereby linking coherence with the fundamental limits of predictability in quantum measurements.

\section{Tripartite QM-EUR and coherence of fermionic field in the context of EGB spacetime \label{GSCDGE}}
\begin{figure}
\begin{minipage}[t]{0.5\linewidth}
\centering
\includegraphics[width=3.0in,height=6.5cm]{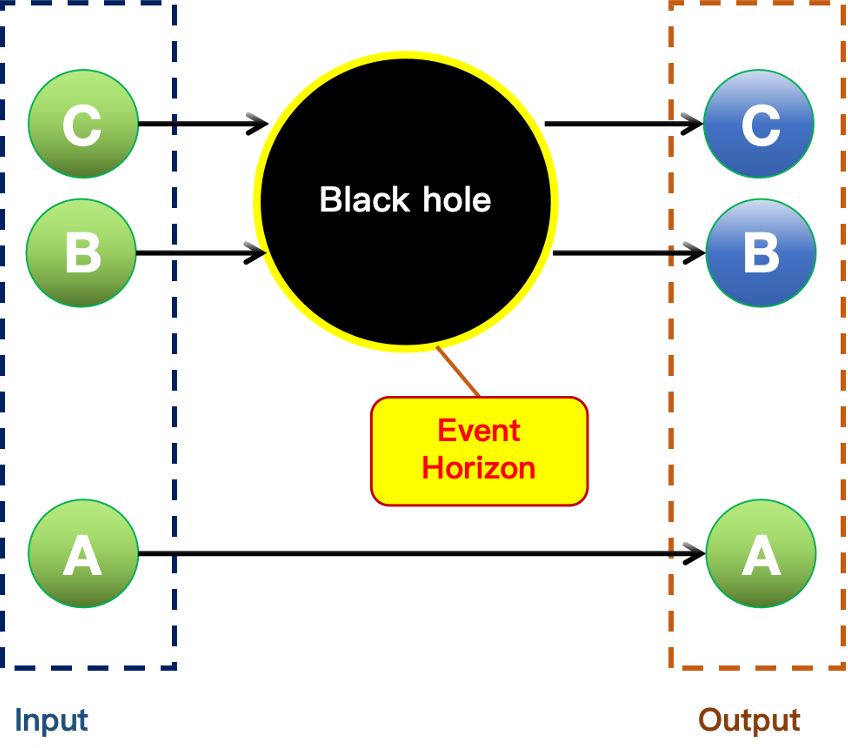}
\label{fig9a}
\end{minipage}%
\caption{This diagram illustrates the configuration where Alice's particle  $A$ remains in the asymptotically flat region, while Bob’s and Charlie’s particles $B$ and $C$ are located near the event horizon of the black hole.}
\label{Fig1}
\end{figure}

\begin{figure}
\begin{minipage}[t]{0.5\linewidth}
\centering
\includegraphics[width=3.0in,height=6.5cm]{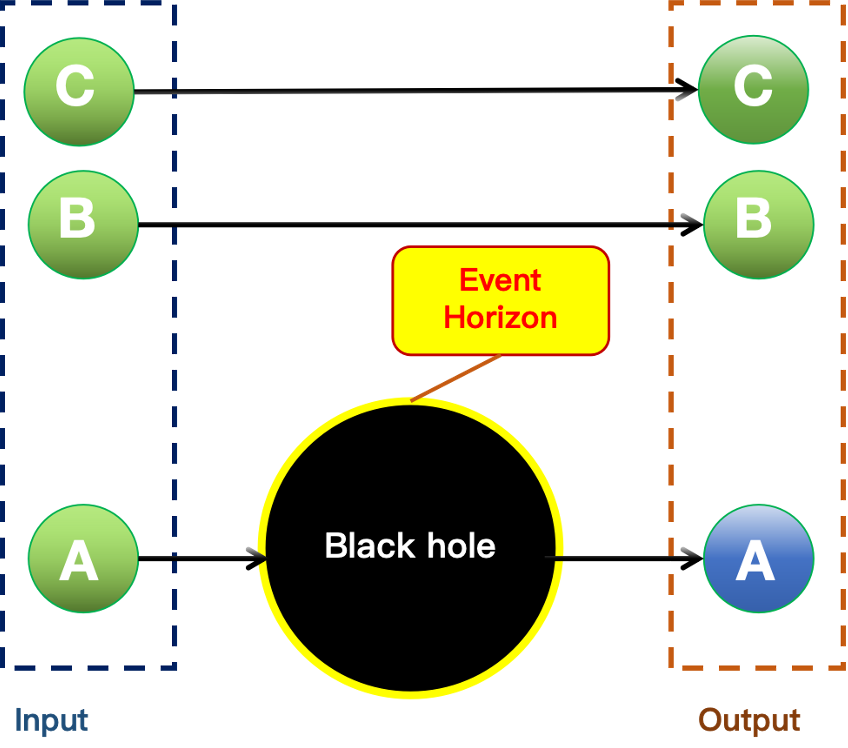}
\label{fig1b}
\end{minipage}%
\caption{This diagram illustrates the scenario in which Alice's particle $A$ is located near the event horizon, whereas Bob's and Charlie's particles $B$ and $C$ remain in the asymptotically flat region.}
\label{Fig2}
\end{figure}

Initially, we consider that Alice, Bob, and Charlie share either a GHZ state
\begin{eqnarray}\label{w31}
\left | GHZ  \right \rangle_{ABC} =\frac{1}{\sqrt{2}} \left ( | 0  \right \rangle_{A}\left | 0  \right \rangle_{B}\left | 0  \right \rangle_{C} +\left | 1  \right \rangle_{A}\left | 1  \right \rangle_{B}\left | 1  \right \rangle_{C}),
\end{eqnarray}
or a W state
\begin{eqnarray}\label{w32}
\left | W \right \rangle_{ABC} =\frac{1}{\sqrt{3} }\left ( \left |1 \right \rangle_{A}  \left |0 \right \rangle _{B} \left |0 \right \rangle_{C}+ \left |0 \right \rangle _{A}\left |1 \right \rangle_{B} \left |0 \right \rangle_{C}+ \left |0 \right \rangle _{A} \left |0 \right \rangle_{B} \left |1 \right \rangle_{C}\right ).
\end{eqnarray}
To investigate tripartite entropic uncertainty and coherence in the EGB spacetime, we analyze two distinct scenarios.  In the first case (Fig.\ref{Fig1}), the parties share a tripartite state $\rho_{ABC}$  prepared in flat Minkowski spacetime. After particle exchange, Alice remains in the asymptotically flat region, while Bob and Charlie freely fall toward the black hole and stay close to the horizon.
Alice then performs either an $X$ or $Z$ measurement on her subsystem and communicates the choice to Bob or Charlie, whose task is to minimize the uncertainty of the corresponding observable. In the second case (Fig.\ref{Fig2}), the measurement protocol is analogous, except that Bob and Charlie remain in the asymptotically flat region, whereas Alice freely falls into the near horizon region of the black hole. In what follows, we compute the tripartite measurement uncertainty and coherence of the GHZ and W states for fermionic field under these two scenarios in the EGB spacetime.

\subsection{GHZ state}
In Scenario 1, Bob and Charlie are located near  the event horizon of the black hole. Using Eqs.(\ref{w29}) and (\ref{w30}), the GHZ state can be rewritten as
\begin{eqnarray}\label{w49}
|GHZ\rangle ^{Case1}_{AB_{\mathrm{I}}B_{\mathrm{II}} C_{\mathrm{I}} C_{\mathrm{II}}}=&&\frac{1}{\sqrt{2}}\big[\cos^{2}\zeta|0\rangle_{A}|0\rangle_{B_{\mathrm{I}}}|0\rangle_{B_{\mathrm{II}}}|0\rangle_{C_{\mathrm{I}}}|0\rangle_{C_{\mathrm{II}}}+\cos\zeta\sin\zeta|0\rangle_{A}|0\rangle_{B_{\mathrm{I}}}|0\rangle_{B_{\mathrm{II}}}\notag\\
&&\otimes |1\rangle_{C_{\mathrm{I}}}|1\rangle_{C_{\mathrm{II}}} +\cos\zeta\sin\zeta|0\rangle_{A}|1\rangle_{B_{\mathrm{I}}}|1\rangle_{B_{\mathrm{II}}}|0\rangle_{C_{\mathrm{I}}}|0\rangle_{C_{\mathrm{II}}}\notag\\
&&+\sin^{2}\zeta|0\rangle_{A}|1\rangle_{B_{\mathrm{I}}}|1\rangle_{B_{\mathrm{II}}} |1\rangle_{C_{\mathrm{I}}}|1\rangle_{C_{\mathrm{II}}}+|1\rangle_{A}|1\rangle_{B_{\mathrm{I}}}|0\rangle_{B_{\mathrm{II}}}|1\rangle_{C_{\mathrm{I}}}|0\rangle_{C_{\mathrm{II}}}\big].
\end{eqnarray}
Since regions I and II are causally disconnected, Bob and Charlie can only access the exterior modes. By tracing out the inaccessible region II, the reduced density matrix $\rho_{AB_{I}C_{I}}$ for the accessible subsystem is obtained as
\begin{eqnarray}\label{w50}	
\rho_{AB_{\mathrm{I}}C_{\mathrm{I}}}^{GHZ}=\frac{1}{2} \begin{pmatrix}
  \cos^{4}\zeta&  0&  0&  0&  0&  0&  0& \cos^{2}\zeta\\
  0&  \cos^{2}\zeta\sin^{2}\zeta&  0&  0&  0&  0&  0& 0\\
  0&  0&  \cos^{2}\zeta\sin^{2}\zeta&  0&  0&  0&  0& 0\\
  0&  0&  0&  \sin^{4}\zeta&  0&  0&  0& 0\\
  0&  0&  0&  0&  0&  0&  0& 0\\
  0&  0&  0&  0&  0&  0&  0& 0\\
  0&  0&  0&  0&  0&  0&  0& 0\\
  \cos^{2}\zeta&  0&  0&  0&  0&  0&  0&1
\end{pmatrix}.
\end{eqnarray}
Employing Eqs.(\ref{w2}) and (\ref{w50}), we can obtain the analytical expression for the measurement uncertainty of the tripartite GHZ state for the fermionic field as
\begin{eqnarray}\label{w51}
U_{A B_{\mathrm{I}} C_{\mathrm{I}}}^{GHZ}=\frac{1}{2}[-\kappa \log_{2}\kappa+\kappa\log_{2}[(2(\kappa+1)]+\log_{2}(2\kappa+2)+\delta],
\end{eqnarray}
where  $\kappa=\sin^{4}\zeta+\sin^{2}\zeta\cos^{2}\zeta$ and $\delta=\cos^{4}\zeta+\sin^{2}\zeta\cos^{2}\zeta$.
According to Eqs.(\ref{w61}) and (\ref{w50}), for Scenario 1, tripartite $l_{1}$-norm of coherence of the GHZ state in the EGB spacetime background is obtained as
\begin{eqnarray}\label{w63}
C_{l_{1}}^{GHZ}(\rho_{AB_{\mathrm{I}}C_{\mathrm{I}}})=\cos^{2}\zeta.
\end{eqnarray}
Similarly, applying Eq.(\ref{w62}) yields the REC
\begin{eqnarray}\label{w64}
C_{REC}^{GHZ}(\rho_{AB_{\mathrm{I}}C_{\mathrm{I}}})&&=-\bigg[ \frac{1}{2}\cos^{4}\zeta\log_{2}({\frac{1}{2}\cos^{4}\zeta})+\frac{1}{2}\sin^{2}\zeta\cos^{2}\zeta\log_{2}(\frac{1}{2}\sin^{2}\zeta\cos^{2}\zeta)\notag\\
&&+\frac{1}{2}\sin^{2}\zeta\cos^{2}\zeta\log_{2}(\frac{1}{2}\sin^{2}\zeta\cos^{2}\zeta)+\frac{1}{2}\sin^{4}\zeta\log_{2}({\frac{1}{2}\sin^{4}})+\frac{1}{2}\log_{2}{\frac{1}{2}}\bigg]\notag\\
&&+\frac{1}{16}(11+4\cos2\zeta+\cos4\zeta)\log_{2}[{\frac{1}{16}(11+4\cos2\zeta+\cos4\zeta)}]\notag\\
&&+\frac{1}{2}\cos^{2}\zeta\sin^{2}\zeta\log_{2}({\frac{1}{2}\cos^{2}\zeta\sin^{2}\zeta})+\frac{1}{2}\cos^{2}\zeta\sin^{2}\zeta\log_{2}({\frac{1}{2}\cos^{2}\zeta\sin^{2}\zeta})\notag\\
&&+\frac{1}{2}\sin^{4}\zeta\log_{2}{(\frac{1}{2}\sin^{4}\zeta)}.
\end{eqnarray}

In Scenario 2, only Alice’s mode is rewritten in terms of the EGB modes, yielding
\begin{eqnarray}\label{w52}
|GHZ\rangle ^{Case2}_{A_{\mathrm{I}} A_{\mathrm{II}} B C}=&&\frac{1}{\sqrt{2}}\big[\cos \zeta|0\rangle_{A_{\mathrm{I}}}|0\rangle_{A_{\mathrm{II}}}|0\rangle_{B}|0\rangle_{C}+\sin \zeta|1\rangle_{A_{\mathrm{I}}}|1\rangle_{A_{\mathrm{II}}}|0\rangle_{B}|0\rangle_{C}\notag\\
&&+|1\rangle_{A_{\mathrm{I}}}|0\rangle_{A_{\mathrm{II}}}|1\rangle_{B}|1\rangle_{C}\big].
\end{eqnarray}
Tracing over the inaccessible mode  $A_{\rm{II}}$ gives the reduced density matrix of the GHZ state
\begin{eqnarray}\label{w53}	
\rho_{A_{\mathrm{I}}BC}^{GHZ}=\frac{1}{2} \begin{pmatrix}
  \cos^{2}\zeta&  0&  0&  0&  0&  0&  0& \cos\zeta\\
  0&  0&  0&  0&  0&  0&  0& 0\\
  0&  0&  0&  0&  0&  0&  0& 0\\
  0&  0&  0&  0&  0&  0&  0& 0\\
  0&  0&  0&  0&  \sin^{2}\zeta &  0&  0& 0\\
  0&  0&  0&  0&  0&  0&  0& 0\\
  0&  0&  0&  0&  0&  0&  0& 0\\
  \cos\zeta&  0&  0&  0&  0&  0&  0&1
\end{pmatrix}.
\end{eqnarray}
From Eqs.(\ref{w2}) and (\ref{w53}) in Scenario 2, the tripartite measurement uncertainty of the GHZ state is obtained as
\begin{eqnarray}\label{w54}
U_{A_{\mathrm{I}} B C}^{GHZ}=1+\frac{1}{2}[-\cos^{2}\zeta \log_{2}(\cos^{2}\zeta)-\sin^{2}\zeta \log_{2}(\sin^{2}\zeta)] .
\end{eqnarray}
Furthermore, the $l_{1}$-norm of coherence reads
\begin{eqnarray}\label{w65}
C_{l_{1}}^{GHZ}(\rho_{A_{\mathrm{I}}BC})=\cos\zeta,
\end{eqnarray}
and REC is given by
\begin{eqnarray}\label{w66}
C_{REC}^{GHZ}(\rho_{A_{\mathrm{\mathrm{I}}}BC})&&=-\bigg[\frac{1}{2}\cos^{2}\zeta\log_{2}({\frac{1}{2}\cos^{2}\zeta})+\frac{1}{2}\sin^{2}\zeta\log_{2}(\frac{1}{2}\sin^{2}\zeta)+\frac{1}{2}\log_{2}{\frac{1}{2}}\bigg]\notag\\
&&+\frac{1}{4}(3+\cos2\zeta)\log_{2}[{\frac{1}{4}(3+\cos2\zeta)}]+\frac{1}{2}\sin^{2}\zeta\log_{2}({\frac{1}{2}\sin^{2}\zeta}).
\end{eqnarray}

\begin{figure}
\begin{minipage}[t]{0.5\linewidth}
\centering
\includegraphics[width=3.0in,height=6.5cm]{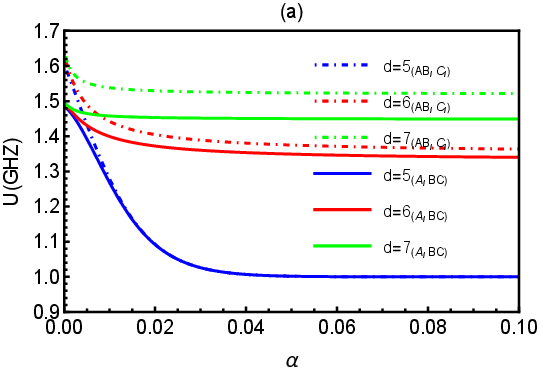}
\label{fig1aa}
\end{minipage}%
\begin{minipage}[t]{0.5\linewidth}
\centering
\includegraphics[width=3.0in,height=6.5cm]{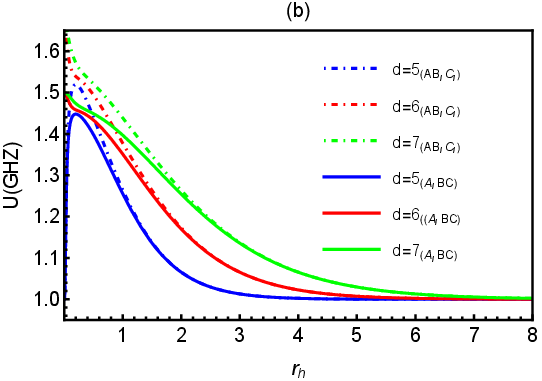}
\label{fig1bb}
\end{minipage}%

\begin{minipage}[t]{0.5\linewidth}
\centering
\includegraphics[width=3.0in,height=6.5cm]{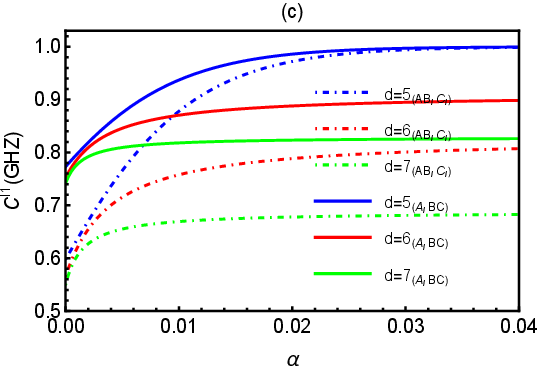}
\label{fig1cc}
\end{minipage}%
\begin{minipage}[t]{0.5\linewidth}
\centering
\includegraphics[width=3.0in,height=6.5cm]{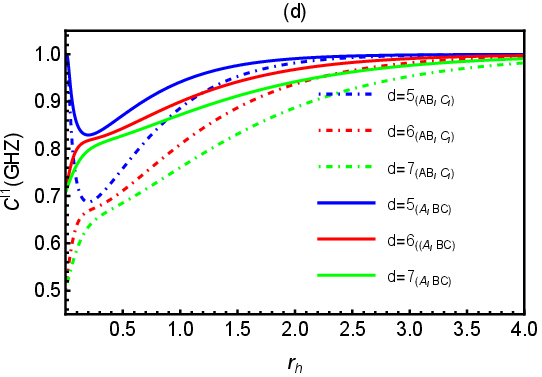}
\label{fig1dd}
\end{minipage}%

\begin{minipage}[t]{0.5\linewidth}
\centering
\includegraphics[width=3.0in,height=6.5cm]{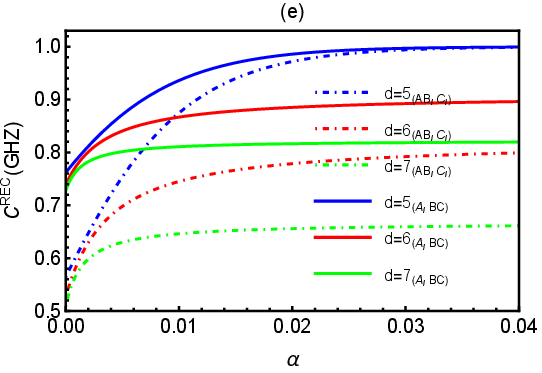}
\label{fig1ee}
\end{minipage}%
\begin{minipage}[t]{0.5\linewidth}
\centering
\includegraphics[width=3.0in,height=6.5cm]{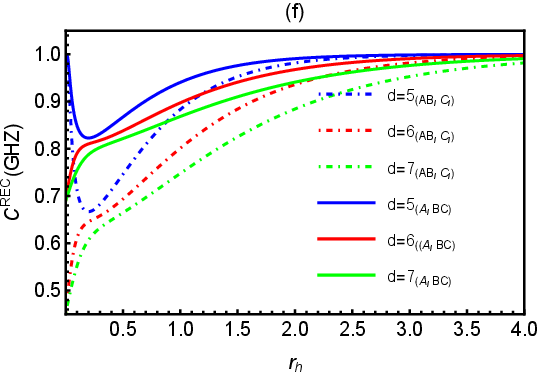}
\label{fig1ff}
\end{minipage}%
\caption{Tripartite measurement uncertainty and quantum coherence of the GHZ state in the fermionic field as functions of the Gauss-Bonnet parameter $\alpha$  and the event horizon radius $r_{h}$ for two distinct physical scenarios.}
\label{Fig3}
\end{figure}

In Fig.\ref{Fig3}, the measurement uncertainty and quantum coherence of the GHZ state in the fermionic field are shown as functions of the Gauss-Bonnet parameter $\alpha$ and the horizon radius $r_{h}$ for two physical scenarios in spacetime dimensions  $d = 5, 6, 7$ with fixed frequency  $\omega=1$. From Fig.\ref{Fig3}(a), we observe that the measurement uncertainty in both scenarios decreases monotonically with increasing $\alpha$. This behavior suggests that in higher-dimensional EGB spacetime, a larger Gauss-Bonnet coupling enhances the rigidity of spacetime, suppresses quantum fluctuations, and consequently reduces measurement uncertainty. Consistently, Fig.\ref{Fig3}(c) and (e) show that both the $l_{1}$-norm of coherence and REC   increase monotonically with  $\alpha$. Fig.\ref{Fig3}(b) shows that for $d = 6, 7$, the uncertainty decreases monotonically with increasing horizon radius $r_{h}$. This can be attributed to the fact that a larger $r_{h}$ corresponds to a lower Hawking temperature $T$, leading to weaker quantum fluctuations and a more rigid spacetime. In contrast, for $d = 5$, the uncertainty exhibits a non-monotonic behavior  first increasing and then decreasing  arising from the peculiar thermodynamic properties of the five-dimensional EGB spacetime as determined by Eqs.(\ref{w14}) and (\ref{w15}). Consistently, Fig.\ref{Fig3}(d) and (f) illustrate that for  $d = 6, 7$, quantum coherence grows with $r_{h}$, since the enlarged horizon suppresses decoherence induced by spacetime curvature and thermal noise. For $d = 5$, however, quantum coherence displays a non-monotonic dependence on $r_{h}$, first decreasing and then increasing, which reflects the distinctive role of the five-dimensional spacetime structure. The differences in the behavior of QMA-EUR and quantum coherence between $d=5$ and $d>5$ dimensions are rooted in the distinct thermodynamic properties of EGB gravity in higher dimensions. Specifically, in $d>5$, the measurement uncertainty decreases monotonically with increasing event horizon radius, while quantum coherence increases, reflecting a more stable thermodynamic state. However, in $d=5$, both the measurement uncertainty and coherence exhibit non-monotonic behavior, which can be attributed to the unique thermodynamic properties of the five-dimensional EGB black hole. Incorporating these observations into the analysis of entropic uncertainty and quantum coherence, we show that the presence of distinctive thermodynamic properties in $d=5$ dimensions leads to non-monotonic changes in these quantities. This highlights the importance of the specific dimensional characteristics of the spacetime when considering quantum informational resources in the context of EGB gravity. This explanation clarifies the reason for the dimensional dependence and provides insight into the interplay between spacetime geometry, quantum fluctuations, and gravitational decoherence, which are crucial for understanding the behavior of quantum resources in curved spacetime.

From Fig.\ref{Fig3}, we find that the overall trends of measurement uncertainty and coherence with respect to the black hole parameters are completely opposite. This contrast arises from the fact that quantum coherence embodies a form of nonclassical correlation: stronger coherence corresponds to reduced measurement uncertainty, and conversely, weaker coherence leads to higher measurement  uncertainty. The opposite variation trends of QMA-EUR and quantum coherence in the EGB spacetime, which both depend on the Gauss-Bonnet coupling and the event horizon radius, can be explained through their relationship with the thermodynamic properties of the black hole in higher dimensions. In higher dimensions, the Gauss-Bonnet coupling $\alpha$ modifies spacetime curvature, influencing thermodynamic quantities such as temperature. An increase in Gauss-Bonnet coupling $\alpha$ or the event horizon radius $r_{h}$ generally enhances thermodynamic stability. This stability suppresses quantum fluctuations, thereby reducing entropic uncertainty. Conversely, greater stability helps preserve quantum correlations, leading to enhanced coherence. Moreover, we also find  that the gravitational effects in Scenario 1 exert a stronger influence on both measurement  uncertainty and coherence of GHZ state compared to those in Scenario 2.

\subsection{W state}
Given that the measurement uncertainty and coherence of the tripartite W state are obtained within the same theoretical framework and through procedures analogous to those used for the GHZ state in the fermionic field, we omit the detailed derivations here for brevity. The complete calculation can be found in the preceding subsection devoted to the GHZ state. For Scenario 1, employing Eqs.(\ref{w29}) and (\ref{w30}), the W state can be rewritten as
\begin{eqnarray}\label{w55}
|W\rangle ^{Case1}_{AB_{\mathrm{I}}B_{\mathrm{II}}C_{\mathrm{I}}C_{\mathrm{II}}}=&&\frac{1}{\sqrt{3}}\big[\cos^{2}\zeta|1\rangle_{A}|0\rangle_{B_{\mathrm{I}}}|0\rangle_{B_{\mathrm{II}}}|0\rangle_{C_{\mathrm{I}}}|0\rangle_{C_{\mathrm{II}}}+\cos\zeta\sin\zeta|1\rangle_{A}|0\rangle_{B_{\mathrm{I}}}|0\rangle_{B_{\mathrm{II}}}|1\rangle_{C_{\mathrm{I}}}|1\rangle_{C_{\mathrm{II}}}\notag\\
&&+\sin\zeta\cos\zeta|1\rangle_{A}|1\rangle_{B_{\mathrm{I}}}|1\rangle_{B_{\mathrm{II}}}|0\rangle_{C_{\mathrm{I}}}|0\rangle_{C_{\mathrm{II}}}+\sin^{2}\zeta|1\rangle_{A}|1\rangle_{B_{\mathrm{I}}}|1\rangle_{B_{\mathrm{II}}}|1\rangle_{C_{\mathrm{I}}}|1\rangle_{C_{\mathrm{II}}}\notag\\
&&+\cos\zeta|0\rangle_{A}|1\rangle_{B_{\mathrm{I}}}|0\rangle_{B_{\mathrm{II}}}|0\rangle_{C_{\mathrm{I}}}|0\rangle_{C_{\mathrm{II}}}+\sin\zeta|0\rangle_{A}|1\rangle_{B_{\mathrm{I}}}|0\rangle_{B_{\mathrm{II}}}|1\rangle_{C_{\mathrm{I}}}|1\rangle_{C_{\mathrm{II}}}\notag\\
&&+\cos\zeta|0\rangle_{A}|0\rangle_{B_{\mathrm{I}}}|0\rangle_{B_{\mathrm{II}}}|1\rangle_{C_{\mathrm{I}}}|0\rangle_{C_{\mathrm{II}}}+\sin\zeta|0\rangle_{A}|1\rangle_{B_{\mathrm{I}}}|1\rangle_{B_{\mathrm{II}}}|1\rangle_{C_{\mathrm{I}}}|0\rangle_{C_{\mathrm{II}}}\big].
\end{eqnarray}

Tracing over the inaccessible modes in region II yields the reduced density matrix
\begin{eqnarray}\label{w56}
	\rho_{AB_{\mathrm{I}}C_{\mathrm{I}}}^{W}=\frac{1}{3} \begin{pmatrix}
  0&  0&  0&  0&  0&  0&  0& 0\\
  0&  \cos^{2}\zeta&  \cos^{2}\zeta&  0&  \cos^{3}\zeta&  0&  0& 0\\
  0&  \cos^{2}\zeta&  \cos^{2}\zeta&  0&  \cos^{3}\zeta&  0&  0& 0\\
  0&  0&  0&  2\sin^{2}\zeta&  0&  \cos\zeta\sin^{2}\zeta&  \cos\zeta\sin^{2}\zeta& 0\\
  0&  \cos^{3}\zeta&  \cos^{3}\zeta&  0&  \cos^{4}\zeta&  0&  0& 0\\
  0&  0&  0&  \cos\zeta\sin^{2}\zeta&  0&   \cos^{2}\zeta\sin^{2}\zeta&  0& 0\\
  0&  0&  0&  \cos\zeta\sin^{2}\zeta&  0&  0&  \cos^{2}\zeta\sin^{2}\zeta& 0\\
  0&  0&  0&  0&  0&  0&  0&\sin^{4}\zeta
\end{pmatrix}.
\end{eqnarray}
From this density matrix, the uncertainty measurement of the W state is obtained as
\begin{eqnarray}\label{w57}
U_{A B_{\mathrm{I}} C_{\mathrm{I}}}^{W}=&&-\frac{\cos^{2}\zeta+2 \sin^{2}\zeta}{3} \log _{2} \frac{\cos^{2}\zeta+2\sin^{2}\zeta}{3}-\frac{\Delta_{-}}{6} \log _{2} \frac{\Delta_{-}}{12}\notag\\
&&+\frac{2\left(\cos^{2}\zeta+\delta\right)}{3} \log _{2} \frac{\cos^{2}\zeta+\delta}{3}-\frac{\Delta_{+}}{6} \log _{2} \frac{\Delta_{+}}{12}\notag\\
&&-\frac{\cos^{2}\zeta}{3} \log _{2} \frac{\cos^{2}\zeta}{3}-\frac{\delta}{3} \log _{2} \frac{\delta}{3}-\frac{\kappa}{3} \log _{2} \frac{\kappa}{3}\notag\\
&&+\frac{\cos^{2}\zeta+2\sin^{2}\zeta+\kappa}{3} \log _{2} \frac{\cos^{2}\zeta+2\sin^{2}\zeta+\kappa}{3},
\end{eqnarray}
where $\Delta_{ \pm}=3 \pm \sqrt{16\sin^{4}\zeta-12 \sin^{2}\zeta+5}$.
Using Eqs.(\ref{w61}), (\ref{w62}), and (\ref{w56}), the $l_{1}$-norm of coherence and REC for the W state in Scenario 1 are found to be
\begin{eqnarray}\label{w67}
C_{l_{1}}^{W}(\rho_{AB_{\mathrm{I}}C_{\mathrm{I}}})=\frac{1}{3}\big(2\cos^{2}\zeta+4\cos^{3}\zeta+4\sin^{2}\zeta\cos\zeta\big),
\end{eqnarray}
\begin{eqnarray}\label{w68}
C_{REC}^{W}(\rho_{AB_{\mathrm{I}}C_{\mathrm{I}}})&&=\frac{1}{24}(1-\cos4\zeta)\log_{2}{\frac{1}{24}(1-\cos4\zeta)}\notag\\
&&+\frac{1}{24}(9-8\cos2\zeta-\cos4\zeta)\log_{2}[{\frac{1}{24}(9-8\cos2\zeta-\cos4\zeta)}]\notag\\
&&+\frac{1}{24}(11+12\cos2\zeta-\cos4\zeta)\log_{2}[{\frac{1}{24}(11+12\cos2\zeta-\cos4\zeta)}]\notag\\
&&+\frac{\sin^{4}\zeta}{3}\log_{2}{\frac{\sin^{4}\zeta}{3}}-\bigg(\frac{\cos^{2}\zeta}{3}\log_{2}{\frac{\cos^{2}\zeta}{3}}\notag\\
&&+\frac{\cos^{2}\zeta}{3}\log_{2}{\frac{\cos^{2}\zeta}{3}}+\frac{2\sin^{2}\zeta}{3}\log_{2}{\frac{2\sin^{2}\zeta}{3}}\notag\\
&&+\frac{\cos^{4}\zeta}{3}\log_{2}{\frac{\cos^{4}\zeta}{3}}+\frac{\cos^{2}\zeta\sin^{2}\zeta}{3}\log_{2}{\frac{\cos^{2}\zeta\sin^{2}\zeta}{3}}\notag\\
&&+\frac{\cos^{2}\zeta\sin^{2}\zeta}{3}\log_{2}{\frac{\cos^{2}\zeta\sin^{2}\zeta}{3}}+\frac{\sin^{4}\zeta}{3}\log_{2}{\frac{\sin^{4}\zeta}{3}}\bigg).
\end{eqnarray}

For Scenario 2, the W state can be expressed as
\begin{eqnarray}\label{w58}
|W\rangle ^{Case2}_{A_{\mathrm{I}}A_{\mathrm{II}}BC}=&&\frac{1}{\sqrt{3}}[|1\rangle_{A_{\mathrm{I}}}|0\rangle_{A_{\mathrm{II}}}|0\rangle_{B}|0\rangle_{C}+\cos \zeta|0\rangle_{A_{\mathrm{I}}}|0\rangle_{A_{\mathrm{II}}}|1\rangle_{B}|0\rangle_{C}\notag\\
&&+\sin \zeta|1\rangle_{A_{\mathrm{I}}}|1\rangle_{A_{\mathrm{II}}}|1\rangle_{B}|0\rangle_{C}+\cos \zeta|0\rangle_{A_{\mathrm{I}}}|0\rangle_{A_{\mathrm{II}}}|0\rangle_{B}|1\rangle_{C}\notag\\
&&+\sin \zeta|1\rangle_{A_{\mathrm{I}}}|1\rangle_{A_{\mathrm{II}}}|0\rangle_{B}|1\rangle_{C}].
\end{eqnarray}
After tracing over the inaccessible mode $A_{\rm{II}}$, the reduced density matrix $\rho_{A_{\mathrm{I}}BC}^{W}$ can be expressed as
\begin{eqnarray}\label{w59}
	\rho_{A_{\mathrm{I}}BC}^{W}=\frac{1}{3} \begin{pmatrix}
  0&  0&  0&  0&  0&  0&  0& 0\\
  0&  \cos^{2}\zeta&  \cos^{2}\zeta&  0&  \cos \zeta&  0&  0& 0\\
  0&  \cos^{2} \zeta&  \cos^{2} \zeta&  0&  \cos \zeta&  0&  0& 0\\
  0&  0&  0&  0&  0&  0&  0& 0\\
  0&  \cos \zeta&  \cos \zeta&  0&  1&  0&  0& 0\\
  0&  0&  0&  0&  0&   \sin^{2}\zeta&  \sin^{2}\zeta& 0\\
  0&  0&  0&  0&  0&  \sin^{2}\zeta&  \sin^{2}\zeta& 0\\
  0&  0&  0&  0&  0&  0&  0&0
\end{pmatrix}.
\end{eqnarray}
From this density matrix, the tripartite measurement uncertainty of the W state in the fermionic field reads
\begin{eqnarray}\label{w60}
U_{A_{\mathrm{I}} B C}^{W}=&& -\frac{2 \cos ^{2}\zeta}{3} \log _{2} \frac{\cos^{2} \zeta}{3}-\frac{\sin^{2}\zeta}{3} \log _{2} \frac{\sin^{2}\zeta}{3} \notag\\
&&-\frac{\Gamma_{-}}{6} \log _{2} \frac{\Gamma_{-}}{12}-\frac{\Gamma_{+}}{6} \log _{2} \frac{\Gamma_{+}}{12} \notag\\
&&+\frac{2}{3} \log _{2} \frac{1}{3}+\frac{4}{3} \log _{2} \frac{2}{3} \notag\\
&&-\frac{1+\sin^{2}\zeta}{3} \log _{2} \frac{1+\sin^{2}\zeta}{3},
\end{eqnarray}
with $\Gamma_{ \pm}=3 \pm \sqrt{4 \cos^{2}\zeta+1}$.
Similarly, the $l_{1}$-norm of coherence of the W state is
\begin{eqnarray}\label{w69}
C_{l_{1}}^{W}(\rho_{A_{\mathrm{I}}BC})=\frac{1}{3}\big(2+4\cos\zeta\big),
\end{eqnarray}
and the REC reads
\begin{eqnarray}\label{w70}
C_{REC}^{W}(\rho_{A_{\mathrm{I}}BC})&&=-\bigg[\frac{\cos^{2}\zeta}{3}\log_{2}{\frac{\cos^{2}\zeta}{3}}+\frac{\cos^{2}\zeta}{3}\log_{2}{\frac{\cos^{2}\zeta}{3}}+\frac{1}{3}\log_{2}{\frac{1}{3}}+\frac{\sin^{2}\zeta}{3}\notag\\
&&\times\log_{2}{\frac{\sin^{2}\zeta}{3}}+\frac{\sin^{2}\zeta}{3}\log_{2}{\frac{\sin^{2}\zeta}{3}}\bigg]+\frac{1}{3}(1-\cos2\zeta)\log_{2}[{\frac{1}{3}(1-\cos2\zeta)}]\notag\\
&&+\frac{1}{3}(2+\cos2\zeta)\log_{2}[{\frac{1}{3}(2+\cos2\zeta})].
\end{eqnarray}

\begin{figure}
\begin{minipage}[t]{0.5\linewidth}
\centering
\includegraphics[width=3.0in,height=6.5cm]{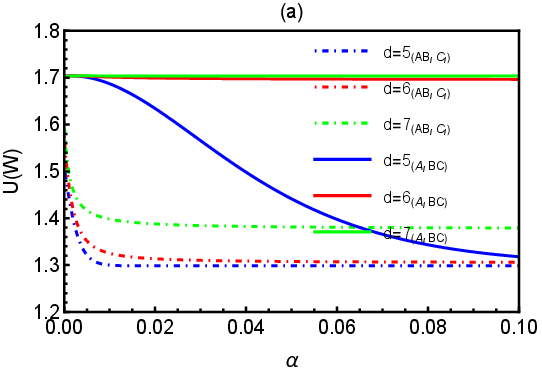}
\label{fig4aa}
\end{minipage}%
\begin{minipage}[t]{0.5\linewidth}
\centering
\includegraphics[width=3.0in,height=6.5cm]{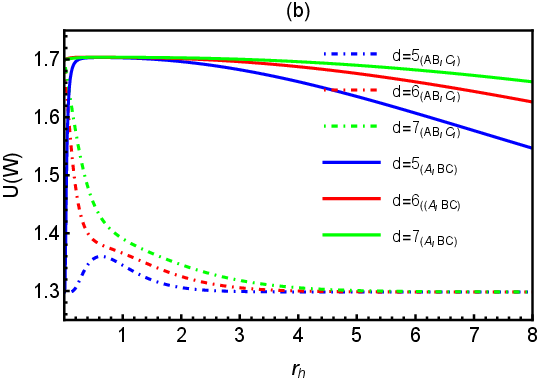}
\label{fig4bb}
\end{minipage}%

\begin{minipage}[t]{0.5\linewidth}
\centering
\includegraphics[width=3.0in,height=6.5cm]{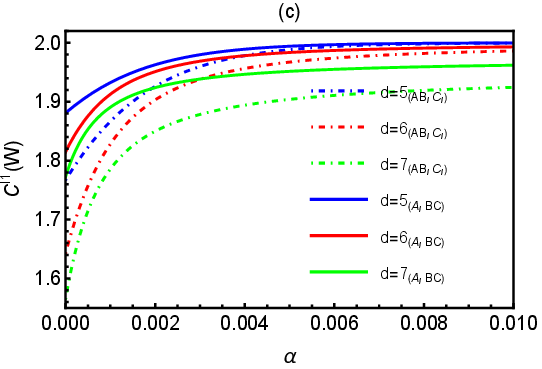}
\label{fig4cc}
\end{minipage}%
\begin{minipage}[t]{0.5\linewidth}
\centering
\includegraphics[width=3.0in,height=6.5cm]{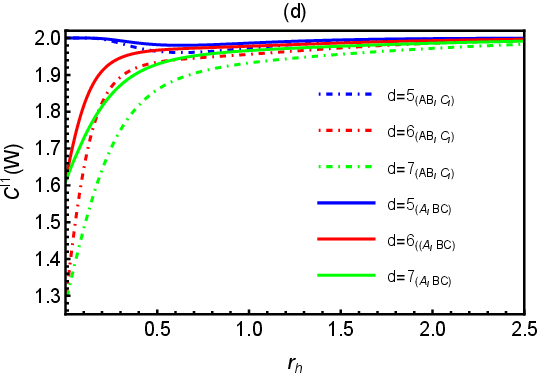}
\label{fig4dd}
\end{minipage}%

\begin{minipage}[t]{0.5\linewidth}
\centering
\includegraphics[width=3.0in,height=6.5cm]{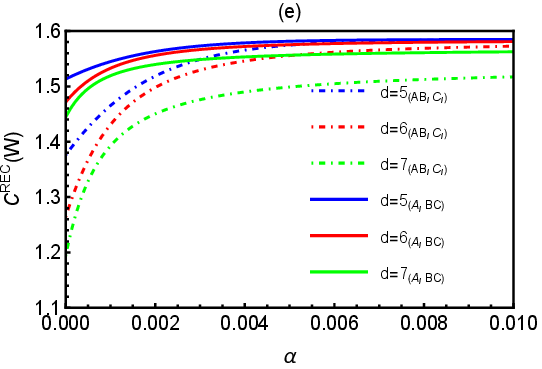}
\label{fig4ee}
\end{minipage}%
\begin{minipage}[t]{0.5\linewidth}
\centering
\includegraphics[width=3.0in,height=6.5cm]{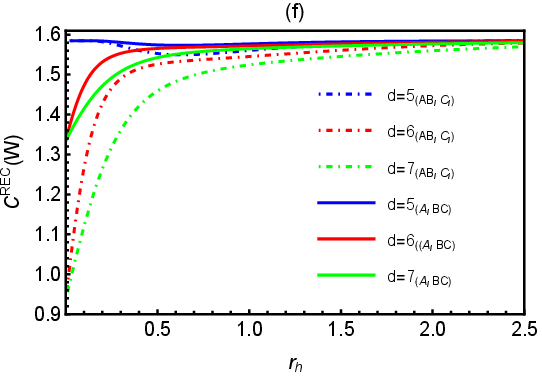}
\label{fig4ff}
\end{minipage}%
\caption{Tripartite measurement uncertainty and quantum coherence of the W state in the fermionic field under two physical scenarios, shown as functions of the Gauss-Bonnet coupling $\alpha$  and the event horizon radius $r_{h}$.}
\label{Fig4}
\end{figure}

In Fig.\ref{Fig4}, we present the tripartite measurement uncertainty, $l_{1}$-norm of coherence, and REC of the W state in the fermionic field as functions of the EGB coupling constant $\alpha$ and the event horizon radius $r_{h}$, for various spacetime dimensions $d$ with fixed frequency $\omega=1$. The behavior of the measurement uncertainty and coherence for the W state exhibits similar qualitative features to those of the GHZ state in the EGB spacetime. This indicates that the Hawking effect increases the measurement uncertainty and reduces quantum coherence, as higher Hawking temperatures correspond to stronger Hawking radiation, which perturbs the quantum system more significantly. For the W state, as the event horizon radius $r_{h}$ increases, the measurement uncertainty in Scenario 1 remains lower than that in Scenario 2. In contrast, for the GHZ state, the situation is reversed: as $r_{h}$ increases, the measurement uncertainty in Scenario 1 is consistently higher than in Scenario 2. Interestingly, regardless of whether the system is in the GHZ or W state, the quantum coherence in Scenario 1 is always lower than in Scenario 2. In both scenarios, under the influence of EGB gravity, the quantum coherence of the W state consistently exceeds that of the GHZ state, indicating that the W state is more advantageous for implementing relativistic quantum information tasks in terms of coherence. However, when considering measurement uncertainty, the GHZ state exhibits stronger resistance to Hawking radiation compared to the W state, thereby demonstrating greater robustness against gravitational decoherence.

\section{ Conclusions  \label{GSCDGE}}
In this study, we have systematically investigated the behavior of tripartite quantum measurement uncertainty and quantum coherence for fermionic field in the context of EGB gravity, focusing on two distinct physical scenarios in a spherically symmetric black hole spacetime. By considering both GHZ and W states as initial quantum states, we derive exact analytical expressions for the measurement uncertainty and coherence measures, including the $l_{1}$-norm of coherence and REC.
Our results reveal that the Gauss-Bonnet coupling constant  $\alpha$ plays a significant role in modulating quantum properties. Specifically, as  $\alpha$ increases, the measurement uncertainty decreases monotonically, while quantum coherence increases monotonically. This indicates that stronger Gauss-Bonnet coupling enhances spacetime rigidity, suppresses quantum fluctuations, and mitigates gravitational decoherence. The behavior with respect to the event horizon radius $r_{h}$ exhibits a dimensional dependence. For spacetime dimensions $(d>5)$, both measurement uncertainty and coherence vary monotonically with $r_{h}$: measurement uncertainty decreases and coherence increases as $r_{h}$ grows, due to lower Hawking temperature and reduced thermal noise. In contrast, for
$(d=5)$,  both quantities show non-monotonic behavior, with measurement uncertainty first increasing and then decreasing, while coherence first decreases and then increases, reflecting the unique thermodynamic stability of a five-dimensional EGB black hole.

Furthermore, we compare the performance of GHZ and W states under gravitational influence. The W state consistently exhibits higher quantum coherence than the GHZ state, making it more suitable for coherence-sensitive relativistic quantum information tasks. Conversely, the GHZ state demonstrates greater robustness against measurement uncertainty induced by Hawking radiation, highlighting its advantage in scenarios where predictability is critical. Interestingly, the two physical scenarios, whether the quantum memory or the measured particle is near the event horizon, lead to qualitatively distinct behaviors. A particularly notable observation is that, regardless of the quantum state used, the quantum coherence in Scenario 1 (quantum memory near the horizon) is consistently lower than in Scenario 2 (measured particle near the horizon). Regarding measurement uncertainty, the two scenarios also exhibit clear differences: for the W state, the measurement uncertainty in Scenario 1 is always lower than in Scenario 2; in contrast, for the GHZ state, the trend is reversed, with the measurement uncertainty in Scenario 1 consistently higher than in Scenario 2.
These results demonstrate that different types of quantum resources exhibit distinct behaviors in curved spacetime. Their characteristics not only determine the stability and robustness of quantum states under strong gravitational fields but also directly influence the efficiency and feasibility of quantum information tasks. Hence, a deeper understanding of these differences provides important guidance for the selection and optimization of quantum states in curved spacetime.

Although the present work is primarily theoretical, recent advances in experimental quantum technologies suggest that the physical scenarios considered here may be explored in realistic or analog settings. Quantum optical experiments using the Micius satellite have already tested quantum field behavior in curved spacetime by transmitting entangled photon pairs through different regions of Earth's gravitational potential, offering a direct experimental platform to investigate gravity induced decoherence and non-unitary effects \cite{qhx7}. Likewise, atomic clock experiments have measured gravitational redshifts within millimeter-scale samples of ultracold strontium atoms, achieving sensitivities relevant for probing quantum coherence at distinct gravitational potentials \cite{qhx8}. In parallel, the development of compact quantum microsatellites and satellite-based quantum key distribution networks shows that quantum coherence and entanglement can be distributed across large gravitational gradients between ground and orbit \cite{qhx9}. Taken together, these developments indicate that analog implementations of relativistic quantum information processes-such as qubits subjected to varying gravitational fields or near an event horizon analog potentials-are increasingly becoming experimentally accessible. The theoretical framework presented in this work therefore provides a foundation for interpreting these emerging experiments and for guiding the design of future analog gravity platforms capable of simulating qubit dynamics and coherence redistribution in curved spacetime.

\begin{acknowledgments}
This work is supported by the National Natural Science Foundation of China (Grant nos. 12175095 and
12205133), the Special Fund for Basic Scientific Research of Provincial Universities in Liaoning under Grant no. LS2024Q002, and LiaoNing Revitalization Talents
Program (XLYC2007047).
\end{acknowledgments}

\appendix
\onecolumngrid
%


\end{document}